\documentclass[preprint,showpacs,preprintnumbers]{revtex4}
\usepackage{amssymb}
\usepackage{graphicx}
\usepackage{dcolumn}
\usepackage{bm}
\usepackage{color}

\input{tcilatex}

\begin{document}

\title{The Deformation Effect on the Electronic Structure of the Graphite
Nanoribbon Arrays}
\author{W. S. Su$^1$, B. R. Wu$^2$\thanks{%
Electronic address: brwu@mail.cgu.edu.tw} and T. C. Leung$^1$}
\affiliation{$^1$Department of Physics, National Chung Cheng Univerisy, Chia-Yi 621,
Taiwan\\
$^2$Center for General Education, Chang Gung University, Tao-Yuan 333, Taiwan}
\date{\today }

\begin{abstract}
We have performed a first-principles study on the deformation effect of the
electronic structures of graphite nanoribbon arrays with zigzag edges on
both sides, and the edge atoms are terminated with hydrogen atoms. A
uniaxial strain is considered to have deformation effect on the graphite
nanoribbons. We found that the antiferromagnetic arrangement of the spin
polarizing edges of graphite nanoribbon is still more favorable than that
with the ferromagnetic arrangement under deformation. We also learned that a
tensile strain increases the magnetization of the graphite ribbon while a
compressive strain decreases it. A positive pressure derivate of the band
gap of antiferromagnetic state is observed for the graphite nanoribbon under
uniaxial strains. The strain changes the shape of band structure and the
band gap; in here, the edge atoms play a crucial role. The deformations are
also found to influence the contribution of the edge atoms to the bands near
the Fermi level. The deformation effect for the graphite nanoribbon under
transverse electric field is also studied.
\end{abstract}

\maketitle






\section{Introduction}

Since graphene had been successfully synthesized in laboratory \cite%
{Ref1,Ref2,Ref3,Ref4,Ref5}, graphene and related structures (e.g. carbon
nanotubes), ignites intense investigation \cite%
{Ref6,Ref7,Ref8,Ref9,Ref10,Ref11}. That is motivated by the fundamental
physics interests and the potentials of versatile applications of the novel
structures formed by graphene. A quasi-one-dimensional graphite nanoribbon
can be constructed by cutting or by patterning a graphene sheet along a
specific direction \cite{Ref4,Ref5}, and it is essentially a strip of
graphene with finite width in nanometer size. Owing to the structure being
similar to carbon nanotubes, the graphite nanoribbons are expected to have
various unique properties and capabilities for the applications of
nanoelectronics and nanomechanical devices \cite%
{Ref5,Ref8,Ref9,Ref10,Ref11,Ref12,Ref13,Ref14,Ref15,Ref16,Ref17,Ref18,Ref19}
.

\mbox{}

The zigzag graphite nanoribbon (ZGNR) is a graphite nanoribbon with zigzag
shaped edges. The ZGNRs have the unusual electronic localized edge states,
which decay exponentially into the center of the ribbon \cite{Ref6}. The
edge states are a twofold degenerate flat band at Fermi level and the flat
band is lasting about one-third Brillouin zone (BZ) starting from the zone
boundary. While counting the spin polarization effect, the spin states are
more favored than the spinless (NM) state. There are two favored spin
states: the ferromagnetic (FM) state and the antiferromagnetic (AF) state.
The AF (FM) state is the configuration with opposite (same) spin orientation
between the two ferromagnetically ordered edges. The ground state of a ZGNR
reveals antiferromagnetic. That is, the two edges of the ribbon prefer to
exhibit opposite spin orientations and the total spin is zero. Owing to the
symmetry breaking and the magnetic interaction between the two edges of a
ZGNR \cite{Ref19}, the degenerated edge states split into two states with
opposite spins and open a gap. The ZGNRs become semiconducting but not
metallic \cite{Ref12,Ref16,Ref17,Ref18,Ref19,Ref20,Ref21}. The band gap and
the magnetization of edge atoms vary with the width of the ZGNR \cite%
{Ref12,Ref19,Ref21}. The ZGNRs may turn to half-metallic when applied an
external electric field across the ribbon that was suggested by Son et al%
\cite{Ref16}. The effect of external electric field makes the band gap
closing in one spin direction and opening in the other spin direction.
Following the previous usage, the gap-opening state referred as $\alpha$%
-spin, and the gap-narrowing state referred as $\beta$-spin. Besides, the
distortions and defects of graphite nanoribbons are also predicted to have
an effect on band structures \cite{Ref15,Ref22,Ref23,Ref24}. In this study,
we mainly present the electronic structures of graphite nanoribbons in the
presence of a uniform uniaxial strain and a transverse electric field.


\section{Calculation Methods}

We consider the deformations of graphite nanoribbons with zigzag edges on
both sides and each of the edge atoms passivated by hydrogen atoms on each
side. The structure is shown in Fig. 1(a), and the coordinate axes are
defined in Fig. 1(b). The graphite nanoribbon arrays consist of
nanographitic strips that are infinitely long in the y direction with the
primitive cells 2.441$\mathring{A}$ for the ZGNRs. This length is the
theoretical result of graphene calculated by density functional theory (DFT) 
\cite{Ref25} and local density approximation \cite{Ref26}. Several
inter-ribbon distances $(D_{x}=3.5\thicksim35\mathring{A})$ are considered
to investigate the inter-layer interaction of the graphite nanoribbon array.
The electric field is applied in the z direction, and the supercell has a
vacuum thickness of 26$\mathring{A}$ in the z direction. A ZGNR is specified
by the number of zigzag chains (N) along the ribbon forming the width and
denote as N-ZGNR, for example, the structure of Fig. 1(a) is referred as a
6-ZGNR. As shown in Fig. 1(a), the width of a ZGNR we defined, here, is the
width without including the hydrogen atoms at the edges. The uniaxial stress
is imposed in the y direction, and strain $(\varepsilon)$ is defined as $%
\varepsilon= \ell-\ell_{0}/\ell_{0}$.

\mbox{}

This work is performed within the DFT and the local spin density
approximation (LSDA) via the Vienna ab initio simulation package (VASP) \cite%
{Ref27,Ref28,Ref29}. The exchange-correlation energy is in the
Ceperley-Alder form \cite{Ref30} and the ultrasoft pseudopotentials \cite%
{Ref28} are used. The plane-wave cutoff up to 358.2 eV has been carried out
and the $1\times 80\times 1$ k-points sampling in the first BZ is used. The
atomic positions are relaxed until the magnitudes of the forces become less
than 0.01 $eV/\mathring{A}$, and the total energies are converged to within
1 meV. In order to simulate a homogenous external electric field (F) in the
z-direction (Fig. 1), a dipole sheet (xy-plane) perpendicular to the ribbon
edge is placed at the center of the vacuum region in the supercell. A dipole
correction is included to avoid artificial Coulomb interactions caused by
the external electric field.


\section{Results and Discussions}

The deformation effect on the ZGNRs with various widths is studied by giving
a uniaxial strain along the y-direction. For the undeformed ZGNRs, we also
got that the AF state is the ground state, and the total energy of the AF
state is lower than that of the FM (NM) state with few (tens) meV per edge
atom. The total energy difference $\Delta E_{FM-AF}$ ($\Delta E_{NM-AF}$)
between the AF and FM (NM) states of a ZGNR lowers (raises) as the
undeformed ZGNR gets wider, it agrees with the results of the previous
studies \cite{Ref12}. For the deformed ZGNRs, the AF state remains the
ground state and the NM state is still less favored than the spin states.
The trend of the $\Delta E_{NM-AF}$ and $\Delta E_{FM-AF}$ varying with the
widths of the ZGNRs is still retained as the ZGNRs are under the same
strain. Fig. 2(a) gives the variations of the $\Delta E_{FM-AF}$ under
various strains. The $\Delta E_{FM-AF}$ raises as the strain increases. It
reveals that the AF state becomes more stable than the FM state for the ZGNR
under a tensile strain. However, the deformation effect on the $\Delta
E_{FM-AF}$ of a ZGNR is less significant when the ZGNR gets wider. Since the
AF state is the ground state of the ZGNRs under strains, here, we only
discuss the deformation effect on the AF spin state. As can be seen in Fig.
2(b), the magnetization of a ZGNR increases as the strain gets large and the
ribbon gets wider. It indicates that the magnetic interaction between the
edge atoms of both sides of ribbon increases under tensile strains, while it
decreases as the ZGRN gets wider. As a result, the increasing (decreasing)
of the magnetic interaction raises (lowers) the $\Delta E_{FM-AF}$ of ZGNRs.

\mbox{}

Figs. 3(a) and 3(b) are corresponding to the variation of band gap of
undeformed ZGRNs and magnetic moment with the width $(w)$, and the band
structures are given in Fig. 3(c). The band gap increases firstly and then
decreases as inversely proportional to the width of N-ZGNR for $N \geq 8$ 
\cite{Ref12}. As the magnetic interaction of the two edges of the ribbon is
one of the reasons to create the band gap of the degenerate edge states, the
strength of the magnetization is related to the variation of the band gap 
\cite{Ref19}. The magnetic interaction is proportional to the square of the
magnetization, while the interaction energy is also inversely proportional
to the width of a ZGNR. Therefore, the band gap of the AF state of a ZGNR
shall be proportional to the square of the magnetization and the inverse of
the width of a ZGNR. The magnetization of the spin state raises
exponentially and saturates when the width of ribbon is greater than 15$%
\mathring{A}$ [Fig.3(b)]. As a result, the variation of the band gap seems
be dominated by the width of ZGRN as the width of ribbon is greater than 15$%
\mathring{A}$ \cite{Ref12}. Though the electron spin density of a ZGNR is
mostly localized around the edge atom, the magnetic interactions of other
atoms near the ribbon edge are not negligible, especially for a short ZGRN $%
(w < 15\mathring{A})$. While counting the contributions of the edge atoms
and other atoms near the ribbon edges to the magnetic interaction, the
calculated band gap can be fitted to the eq. (1) below: 
\begin{eqnarray}
E_{g}=479.07{\frac{m^{2}_{1}}{{w+36.80}}}+913.11{\frac{m^{2}_{2}}{{w-1.81}}}
\end{eqnarray}
Where $m_{1}$ and $m_{2}$ are corresponding to the magnet moments of edge
atom, and other atoms near the ribbon edge, and $w$ is in $\mathring{A}$.
Similar to the ribbon without deformation, the band gap of a wide ZGNR under
a fixed strain also shows inversely proportional to the width of the ZGNR.
As shown in Fig.3(c), the ZGNRs have flat conduction edge band and the slope
of the valence edge band shows no significant change expect for 2-ZGNR. The
magnetic interaction takes the conduction and valence edge bands apart and
makes little change in the shape of the edge bands except for that of 2-ZGNR.

\mbox{}

Fig. 4(a) gives the band gaps of the AF states with various strains. We find
a positive pressure derivative of band gap of ZGRNs. The band gap of a ZGNR
under uniaxial strains shows an interesting behavior. The band gap of a ZGNR
gets opening under compressive strains $(\varepsilon<0)$, and reaches a
maximum at a critical compressive strain $\varepsilon_{c}$; then turns to
get closing as the ZGNR is continuously compressed. On the contrary, the
band gap of a ZGNR gets closing under tensile strains $(\varepsilon>0)$. The
band gap of a ZGNR becomes continuously opening and reaches to a minimum at
a critical tensile strain $\varepsilon_{T}$, then gets opening while the
ribbon is continuously elongated. As can be seen in Fig. 4(a), the critical
compressive (tensile) strain $\varepsilon_{c}(\varepsilon_{T})$ are related
to the width of a ZGNR, $\varepsilon_{c}$ decreases and $\varepsilon_{T}$
increase, as the ribbon gets wider. The strain induced band gap variation of
a wide ribbon is stronger than that of a narrow one. In Fig. 4(b), the edge
bands of 10-ZGRN under strains are plotted. We found that band structure of
a ZGNR is sensitive to the strain. It is notable the uniaxial strain changes
the shape of the band structure of a ZGNR, especially the conduction edge
band. The band bending of the edge states dominates the behavior of band gap
of the deformed ZGNR. In additional to the factor of band bending, the
magnetic interaction is the other factor to affect the behavior of band gap.
For an elongated ZGNR, the band bending tends to open the band gap, while
the magnetic interaction to close. The competition of the two factors
determined the behavior of the band gap of a ZGNR under strains.

\mbox{}

For a deformed ZGNR, a tensile strain increases the magnetization of the
edge atoms [Fig. 2(b)] but decreases the width of the ribbon; while a
compressive strain gives oppositely results. Following the rule of the
magnetic interaction to the band gap of a ZGNR state above, a ZGNR shall
have a negative pressure coefficient of band gap. However, the band gap of a
ZGNR has a positive coefficient rather than a negative one for $\varepsilon
_{c}<\varepsilon <\varepsilon _{T}$; whereas, it has a negative pressure
coefficient for $\varepsilon <\varepsilon _{c}$ and $\varepsilon
_{T}<\varepsilon $. As can be seen in Fig. 4(b), the gap of the edge states
at $k_{F}$ (Here, the wave vector of the top of the valence band is referred
as $k_{F}$) and BZ boundary gets apart under a tensile strain, while the gap
gets closing under a compressive strain. Thus, a stronger magnetic
interaction corresponds to a larger gap. It reveals the magnetic interaction
actually affects the edge bands of ZGNRs under strains. When strain is in
the region of $\varepsilon _{c}<\varepsilon <\varepsilon _{T}$, a ZGNR has a
positive pressure derivative of band gaps. Similarly, the band gap of
diamond also exhibits a positive pressure derivative. However, the mechanism
of both cases is different. The opening of the band gap of diamond is due to
the apart of conduction and valence bands, while the variation of band gaps
of ZGNR is caused by band bending of the conduction edge band. Strain makes
the edge bands bending and the band bending of conduction edge band
determinates the variation of band gap. Compressive strains lowered the
conduction edge band near $k_{F}$ point and lifted the conduction edge band
near the BZ boundary. The minimum of the conduction edge band is at the BZ
boundary as $\varepsilon >\varepsilon _{c}$, while shift to $k_{F}$ at the
critical compressive strain of $\varepsilon _{c}$. For a ZGNR under a
compressive strain, the minimum of conduction edge band is lifted and causes
the gap opening. As strain is in the region of $\varepsilon <\varepsilon
_{c} $ or $\varepsilon >\varepsilon _{T}$, the effect of magnetic
interaction exceeds that of band bending. Thus, the band gap of a ZGNR
reaches a maximum at the critical compressive strain $(\varepsilon _{c})$.

\mbox{}

As shown in Fig. 4(b), the edge bands of an undeformed 10-ZGNR is lasting
around one-fourth BZ starting from the BZ boundary, and the top of the
valence band locates at $k_{F}= 0.75 \pi/a_{y}$. As the strain increases,
the $k_{F}$ shifts to low wave vector and the edge bands extend to the low $%
k $ region. Moreover, the edge band extends more than one-third of BZ as the
strain is up to 0.12. The edge bands of the ZGNR under a tensile strain are
longer and steeper than the ZGNR under a compressive strain. It is found
that tensile strain reduces the contribution of the edge atoms to the edge
bands near the BZ boundary but raises that for $k_{F}<k<0.9 \pi/a_{y}$. It
is contrary while ZGNRs is under compressive strains. As for the edge band
of $\alpha$-spin state, the valence edge band is mainly contributed by the
edge atoms at one side of the ribbon, while the conduction edge band by the
edge atoms at the other side. In the same manner, the edge band of $\beta$%
-spin is mainly contributed by the edge atoms at opposite side. When the
ribbon is under tensile strains, the edge bands near the BZ boundary are
both lowered, while the conduction edge band near $k_{F}$ point is lifted.
Thus, both the edge bands become steeper when the ribbon is under tensile
strains. However, the conduction edge band near the BZ boundary turns to be
lifted as $\varepsilon>\varepsilon_{T}$. As a result, the band gap of the
ribbon becomes opening for $\varepsilon>\varepsilon_{T}$. However, the
strain changes the shape of the edge band of the ZGNRs. The variation of
band gap is dominated by the band bending of the edge states for $%
\varepsilon_{c}<\varepsilon<\varepsilon_{T}$, while dominated by the
magnetic interaction for the strain in the region of $\varepsilon<%
\varepsilon_{c}$ and $\varepsilon>\varepsilon_{T}$.

\mbox{}

With applied transverse electric fields, the band gap of a ZGNR related to $%
\alpha$-spin get wide, while the band gaps related to $\beta$-spin get
narrow. Similar to the previous study of a ZGNR under external field, the
reduction and increasing of the band gap strongly depends on the width of
the ZGNR. When applied the same strength of electric field, a wide ZGNR is
easier to reach the half-metallic state than the narrow one. That is owing
to the reduction of the band gap depends on the voltage between the two
edges, and the voltage is proportional to the width of the ZGNR. We also
investigate the deformed ZGNR under transverse external electric fields. The
variation of band gap of the deformed 10-ZGNR with external electric fields
is shown in Fig. 5(a) and the band structures of the undeformed 10-ZGNR with
external electric fields are given in Fig. 5(b). With external electric
field, the band gaps of deformed ZGNRs are all reduced no matter how the
strain is compressive or tensile. When applied a fixed external electric
field, the behavior of band gaps of a ZGNR with various strains is similar
to that without external electric field. Besides, the value of the critical
compressive strain $\varepsilon_{c}$ increases as the strength of the
external field increases, as can be seen in Fig. 5(a). Because the external
electric field and the compressive strain both title the conduction edge
band, but in opposite directions [see Fig. 5(b) and Fig. 4(b)]. It needs
larger strain to against the effect caused by electric field, as a result,
the critical compressive strain increases under a transverse electric field.


\section{Conclusions}

Conclusively, we have performed a detailed investigation of the ZGNRs under
uniaxial strains and under external transverse electric fields. The
antiferromagnetic state of a ZGNR remains the ground state under
deformations. A tensile strain increases the magnetization of the ZGNR,
while a compressive strain decreases it. An interesting behavior of band
gaps of the ZGNRs under strains is found. The strain changes the shape of
the band structure and the band bending of conduction edge band dominate the
variation of band gap for $\varepsilon_{c}<\varepsilon<\varepsilon_{T}$. The
band gap of the AF state of a deformed ZGRN is also reduced with an external
electric field. As the applied an external electric filed is fixed, the
behavior of the band gap with various strains is similar to that without
external electric fields.


\section{Acknowledgement}

We would like to acknowledge the National Center for Theoretical Sciences
(NCTS) in Taiwan and the financial support from National Science Council
(NSC) of Taiwan under Grant Nos. NSC96-2738-M-002-003 and
NSC96-2112-M-194-012-MY3, and from a grant of computer time at the National
Center for High Performance Computing in Taiwan.

\clearpage

\begin{figure}[tbp]
\begin{flushleft}
\section{Figure Captions}
\end{flushleft}
\includegraphics[width=0.5\textwidth,angle=0]{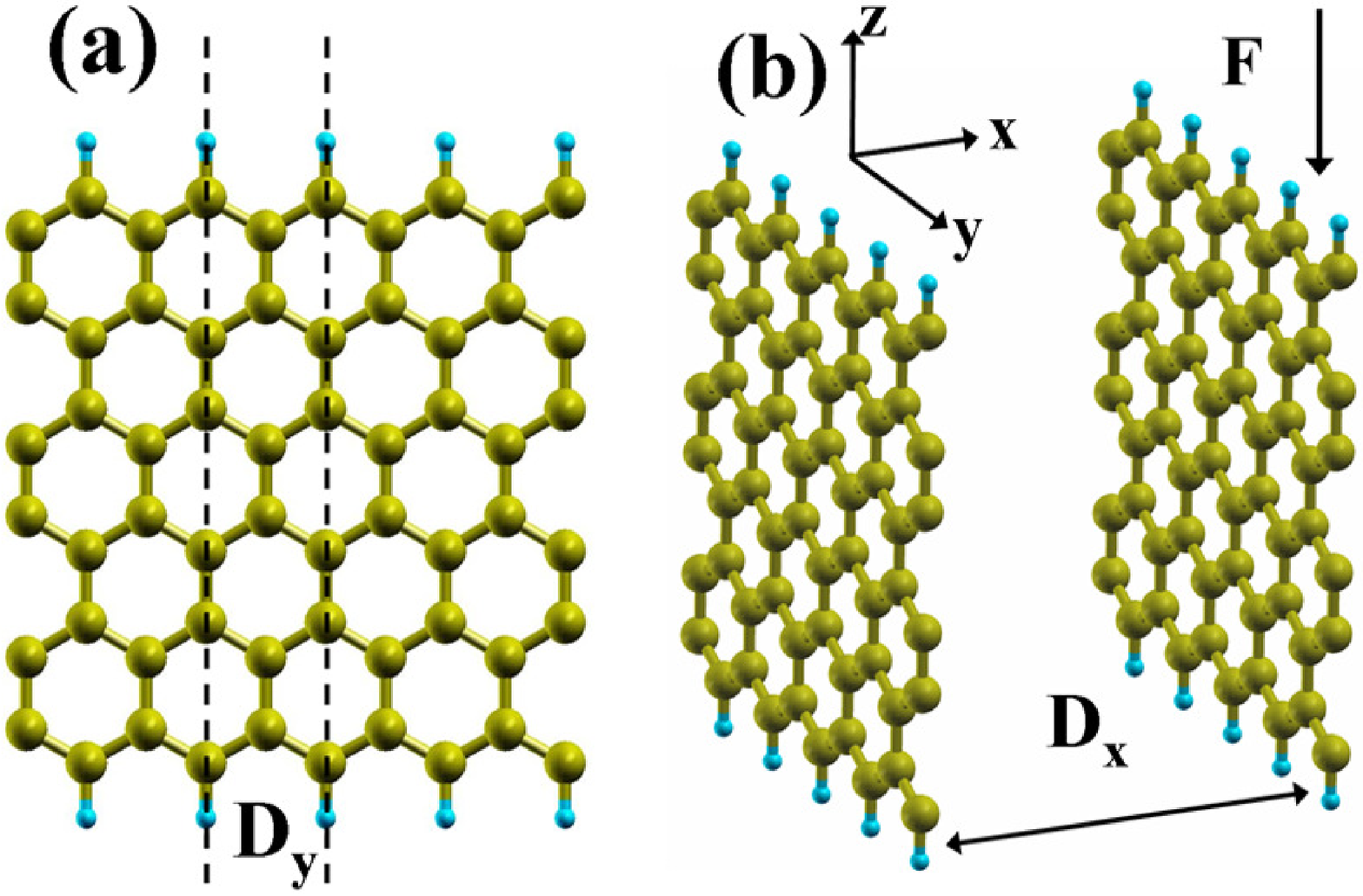} 
\caption{(Color online) (a) Structure of the 6-ZGNR. The boundary of the
primitive cell in the y direction is shown with the dashed lines. (b)
Orientation of the unit cell for the graphite nanoribbons with edges
terminated by H atoms (small blue circles). The external electric field F is
represented by an arrow.}
\label{fig:1}
\end{figure}

\begin{figure}[tbp]
\includegraphics[width=0.5\textwidth,angle=0]{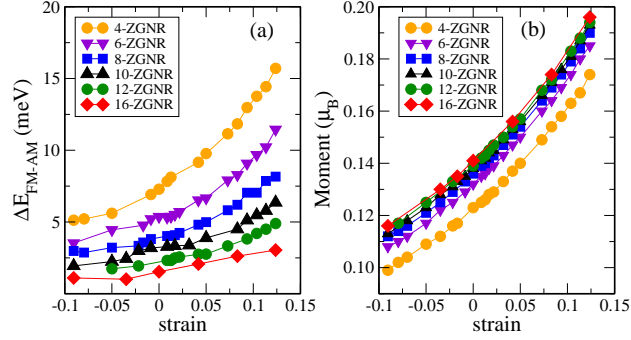} 
\caption{(Color online) (a) Total energy difference between AF state and FM
state $(\Delta E_{FM-AF}=E_{FM}-E_{AF})$ as a function of strain with
various lengths. (b) Magnetic moment of AF state as a function of strain
with various lengths.}
\label{fig:2}
\end{figure}

\begin{figure}[tbp]
\includegraphics[width=0.5\textwidth,angle=0]{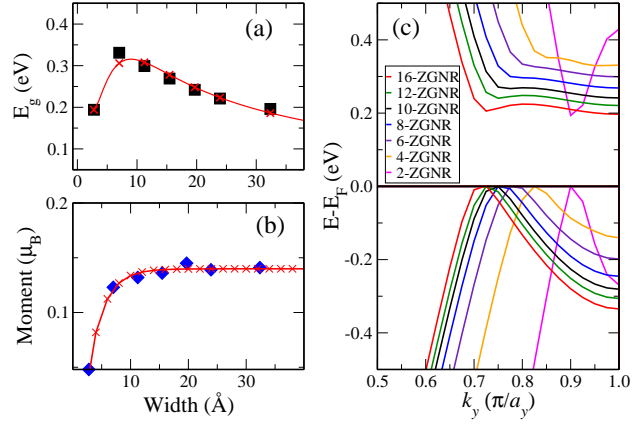} 
\caption{(Color online) (a) Energy gap as a function of width of ZGNR. (b)
Magnetic moment of AF state as a function of width of ribbon. (c) The edge
band of the undeformed ZGNRs with various widths.}
\label{fig:3}
\end{figure}

\begin{figure}[tbp]
\includegraphics[width=0.5\textwidth,angle=0]{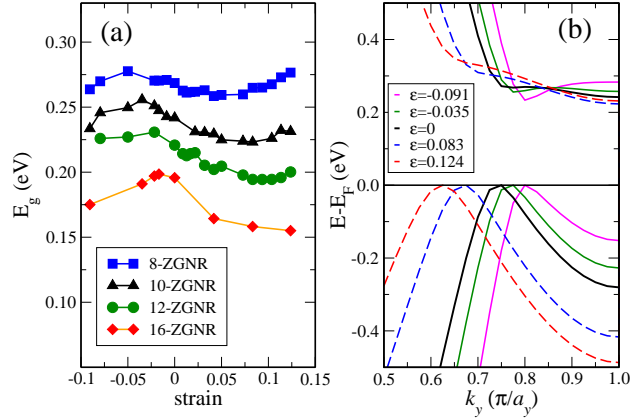} 
\caption{(Color online) (a) Energy gaps of the ZGNRs as functions of strain
with various widths. (b) The edge band of the 10-ZGNR under various strains.
Solid lines and dashed lines represent the band structure with compressive
and tensile strain, respectively.}
\label{fig:4}
\end{figure}

\begin{figure}[tbp]
\includegraphics[width=0.5\textwidth,angle=0]{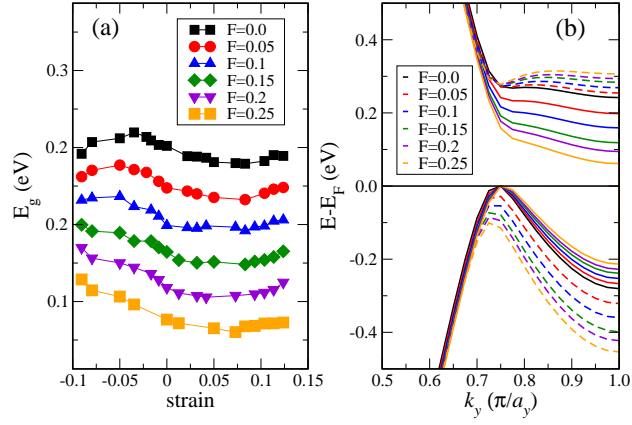} 
\caption{(Color online) (a) The band gaps of the 10-ZGNR with external
electric field as a function of strain. (b) The band structure of an
undeformed 10-ZGNR under external electric field. The dashed lines and solid
lines are shown as $\protect\alpha $-, and $\protect\beta $-spin states. The
line of colors in black, red, green, blue, cyan and pink denote the external
electric field F=0, 0.05, 0.1, 0.15, 0.2, and 0.25 $V/\mathring{A}$,
respectively.}
\label{fig:5}
\end{figure}

\end{document}